\documentclass[12pt]{article}

\usepackage{epsfig}

\newcommand{\bea}{\begin{eqnarray}}
\newcommand{\eea}{\end{eqnarray}}
\newcommand{\be}{\begin{equation}}
\newcommand{\ee}{\end{equation}}

\newcommand{\as}{\alpha_s}

\newcommand{\ar}{a_s}

\title{Gottfried sum rule in QCD NS analysis of DIS fixed target data
\author{A.V.~Kotikov,
V.G.~Krivokhizhin,
B.G.~Shaikhatdenov\\
Joint Institute for Nuclear Research, 141980, Dubna, Russia } }

\begin{document}
\maketitle
\abstract{Deep inelastic scattering data on $F_2$ structure function obtained 
in the fixed-target experiments were analysed in the valence quark
approximation with a next-to-next-to-leading-order accuracy. Parton distribution
functions are parametrized by using information from the Gottfried sum rule. 
The strong coupling constant is found to be $\as(M_Z^2) = 0.1180 \pm 0.0020~\mbox{\scriptsize{(total exp.error)}}$, which coincides very well with the average world value $\as^{\rm PDG}(M_Z^2) = 0.1181 \pm 0.0011$ updated recently in a PDG report.
The result for the second moment of the difference in $u$ and $d$ quark distributions 
$<\!\!x\!\!>_{u-d}=0.187 \pm 0.021$ is seen to be well compatible with the latest LATTICE result
$<\!\!x\!\!>_{u-d}^{\rm LATTICE}=0.208 \pm 0.024$.
} \\

$PACS:~~12.38~Aw,\,Bx,\,Qk$\\


\section{ Introduction }

The deep-inelastic lepton--hadron scattering (DIS) is a basic process for studying 
the properties of parton distribution functions (PDFs), which are universal 
in the considered scheme of calculations and can be used in the subsequent analyses of various processes.

The valence quark approximation\footnote{Often it is called a nonsinglet (NS) approximation and so will we do.} is a simplest one because of the absence of gluons. Due to the QCD factorization and evolution, there is significant correlation between the values of the strong coupling constant $\as(M_Z^2)$ and
gluon density. That is why NS approximation is very well suited for evaluating the strong coupling constant, though it is of course cannot be considered complete.

The modern level of approximation is the next-to-next-to-leading order (NNLO) one.
First NS analyses at the NNLO level~\cite{PKK} showed a small decrease in the average values of $\as(M_Z^2)$  with respect to the corresponding next-to-leading order (NLO) values $\as^{\rm NLO} (M_Z^2)$. Similar behavior has been found also in other analyses (see, for example,~\cite{Accardi:2016ndt} for a recent review). 

We note that the NNLO analyses in~\cite{PKK} contained only BCDMS data, which as a rule led to somewhat low values of the strong coupling constant. Sometimes it is called a BCDMS effect.
As it was recently shown in~\cite{Kotikov:2014cua,Kotikov:2015zda}
the BCDMS data~\cite{BCDMS1,BCDMS2,BCDMS3} can be responsible for the
large differences in both the cross-section values and extracted parameters observed in the analyses
done by using Alekhin--Blumlein--Moch (ABM)~\cite{ABM1} and Jimenez-Delgado--Reya (JR) PDF sets~\cite{JR} on one hand and CTEQ~\cite{CT10}, NN21~\cite{NN21} and  MSTW~\cite{MSTW} ones on the other.
\footnote{The low values of $\as(M_Z^2)$ in~\cite{ABM1,JR} 
can partially be explained~\cite{Thorne:2014toa} by the usage of the fixed flavor number scheme in deriving the ABM sets.}
Indeed, the results in~\cite{ABM1,JR} were obtained by fitting mostly DIS data, while other groups~\cite{CT10,NN21,MSTW} included in their fits some other experimental data (see~\cite{Watt:2011kp} 
and references therein).

In~\cite{Kri2} it was shown that those precise BCDMS data were collected with large systematic 
errors within certain ranges, which can presumably be responsible for an effective decrease in the value of
$\as(M_Z^2)$ (see~\cite{Kri2,KK2001,KKPS,Kotikov:2014cua,Kotikov:2015zda} and~\cite{KK2009}).

The purpose of the present paper is to constrain PDFs by using the Gottfried sum rule~\cite{Gottfried:1967kk} in the analyses and to evaluate the 
strong coupling constant and the second Mellin moment $<\!\!x\!\!>_{u-d}$ in that case. The latter is a point of interest for the lattice QCD community (among other quantities of course); therefore, we compare it with their numbers.

Note that with no usage of the Gottfried sum rule the results of the fits done 
in our previous paper \cite{Kotikov:2014cua} lead to a very large error while estimating 
the value of the second Mellin moment. Here we will incorporate that sum rule
explicitly into the parameterization of parton densities (see subsection 2.1 
below) in order to reduce that error.
  
\section{ Approach }
As we have already mentioned in the Introduction,
one of the most accurate processes to extract $\as(M_Z^2)$ is the valence part
of DIS structure function (SF) $F_2$, which is free of any correlations with the gluon density,
hence the consideration limited to the valence part only.

Since we restrict analysis to the large $x$ region, the data on the total 
structure function $F_2(x,Q^2)$ can be considered in the nonsinglet approximation.
Our study follows the original analysis of BCDMS Collaboration 
done in \cite{Benvenuti:1989rg}, where it was shown that the nonsinglet approximation is valid 
at large $x$ values (see also some previous studies of gluon densities in 
\cite{Benvenuti:1987zm}) in the study  of the slope of $F_2$, $d\ln F_2(x,Q^2)/d\ln Q^2$.

Theoretically, it can be explaned as follows. 
The study of $Q^2$-evolution of SF $F_2$ with the purpose of extracting
the (normalization of) the strong coupling constant is based upon the
Dokshitzer-Gribov-Lipatov-Altarelli-Parisi (DGLAP) equations \cite{DGLAP},
where PDF slopes $d\ln f_a(x,Q^2)/d\ln Q^2$ $(a=q,g)$ are related
to the integrals of parton densities themselves, which are taken into account
together with the corresponding splitting functions $P_{ab}(x)$ $(a,b=q,g)$.
The leading contribution to the SF $F_2$ comes from the quark density 
$f_q(x,Q^2)$ while the gluon distribution starts to contribute only at the 
next-to-leading-order.

For large $x$ values, the gluon density is  significantly softer as 
compared to the quark one, $f_g(x,Q^2)/ f_q(x,Q^2) \sim (1-x)$, which is
in agreement with the predictions from the quark counting rules~\cite{Matveev:1973ra}.
Moreover, it boils down to evaluating the slope 
of quark distribution $d\ln f_q(x,Q^2)/d\ln Q^2$ together with 
the splitting function $P_{qg}(x)$, which produces an additional suppression 
$\sim (1-x)$ at large $x$. It is of course a minor contribution as 
compared to the corresponding quark ones, with the latter being not so soft 
and contributing along with the splitting function $P_{qq}(x)\sim \ln(1-x)$ 
at large $x$ values. 

Beyond the leading order of perturbation theory (PT)
the gluon density also contributes directly to the SF $F_2$, but its
contribution is suppressed  by the factor $\sim \alpha_s$ as compared to the quark one.
Even more, a relative contribution of the gluon density is additionally suppressed.
Indeed, note that the quark density contribution to the SF $F_2$ is multiplied by
the coefficient function  $\sim \alpha_s \ln^2 (1-x)$ and $\sim \alpha_s^2 \ln^4 (1-x)$ \cite{vanNeerven:1999ca}
in NLO and NNLO, respectively, while that in front of the gluon density beyond LO
has the following form: $\sim (1-x) \alpha_s \ln (1-x)$ and $\sim (1-x)\alpha_s^2 \ln^3 (1-x)$
 (see~\cite{vanNeerven:2000uj}).
These latter asymptotic expressions make one to expect that relative contributions
of the gluon density at NLO and NNLO levels of accuracy are somehow on par,
and this gives some credibility to the observed consistency in the results 
for the strong coupling constant values obtained within an NS approach 
and in the complete analysis both carried out in NLO approximation (see~\cite{Benvenuti:1989rg,Benvenuti:1987zm,KK2001}).
Therefore, the same level of consistency could be expected also in the NNLO approximation.
Nevertheless, such kind of speculative considerations should be verified
by relevant numerical studies. We plan to perform them by using PDFs generated
from the codes created and nurtured by the groups~\cite{ABM1}--\cite{MSTW}.
 
DIS structure function $F_2(x,Q^2)$ is dealt with by analyzing SLAC, NMC and BCDMS 
experimental data~\cite{BCDMS1,BCDMS2,BCDMS3,SLAC1,SLAC2,NMC} at NNLO of massless perturbative QCD.
As in our previous papers the function $F_2(x,Q^2)$ is represented as a sum of the leading twist 
$F_2^{\rm pQCD}(x,Q^2)$ and twist four terms
\be
F_2(x,Q^2)=F_2^{\rm pQCD}(x,Q^2)\left(1+\frac{\tilde h_4(x)}{Q^2}\right)\,,
\label{1.1}
\ee
where $F_2^{\rm pQCD}(x,Q^2)$ denotes the twist-2 part with target mass corrections included.
The second term $\sim \tilde h_4(x)$ denotes nonzero twist-4 corrections.
For more details concerning an approach to analysing the experimental data we adopt refer to~\cite{KK2001,KK2009}.

\subsection{Parton densities}
The moments ${\bf f}_i(n,Q^2)\,, (i = \mbox{ns, q, g})$ at some $Q^2_0$ is a theoretical input to the analysis. In the fits of data with the cut $x\geq 0.25$ imposed, only the nonsinglet parton
density is in the game and the following patametrization at the normalization point is used
for the deutron $\tilde{\bf f}_{\rm{D}}(x,Q^2)$, carbon $\tilde{\bf f}_{\rm{C}}(x,Q^2)$ and proton
  $\tilde{\bf f}_{\rm{H}}(x,Q^2)$ cases:
\bea
&&{\bf f}(n,Q^2) = \int_0^1 dx x^{n-2} \tilde{\bf f}(x,Q^2),~~~ \nonumber \\
&&\tilde{\bf f}_{\rm{D,C}}(x,Q^2) = A_{\rm{D,C}}(Q^2) [1-x]^{b_{\rm{D,C}}(Q^2)}[1+d_{\rm{D,C}}(Q^2)x]\,, ~~~ \nonumber \\
&&\tilde{\bf f}_{\rm{H}}(x,Q^2) = \tilde{\bf f}_{\rm{D}}(x,Q^2) + \frac{I_2}{N_{\rm{H}}}
x^{\lambda_{\rm{H}}(Q^2)} [1-x]^{b_{\rm{H}}(Q^2)}[1+d_{\rm{H}}(Q^2)x]\,,
\label{4}
\eea
where $A(Q^2)$, $\lambda(Q^2)$, $b(Q^2)$ and $d(Q^2)$ are some coefficient functions.
Here, the normalization
\bea
&&N_{\rm{H}} =  \int_0^1 \frac{dx}{x} x^{\lambda_{\rm{H}}(Q^2)} [1-x]^{b_{\rm{H}}(Q^2)}[1+d_{\rm{H}}(Q^2)x] \nonumber \\
&&= \frac{\Gamma(\lambda_{\rm{H}}(Q^2))\Gamma(1+b_{\rm{H}}(Q^2))}{\Gamma(1+\lambda_{\rm{H}}+b_{\rm{H}}(Q^2))}
\left[1+ d_{\rm{H}}(Q^2)\frac{\lambda_{\rm{H}}(Q^2)}{1+\lambda_{\rm{H}}+b_{\rm{H}}(Q^2)}\right]
\label{4.1}
\eea
and the factor $I_2$ is related with the Gottfried sum rule $I_G$ \cite{Gottfried:1967kk} (see also the review~\cite{Kumano:1997cy}) as follows
\be
I_G(Q^2)=C_{\rm{NS}}(a_s(Q^2))I_2(Q^2)=\left(1+B_G
a_s^2(Q^2)\right)I_2(Q^2),~~ a_s(Q^2)= \frac{\alpha_s(Q^2)}{4\pi},
\label{I_G}
\ee
where~\cite{Ross:1978xk,Kataev:2003en,Kotikov:2005gr}
\be
B_G \equiv B_{\rm{NS}}^{(2)}(n=1) \approx -0.615732 \, .
\label{B}
\ee

As can be seen in Eq. (\ref{4}), the Gottfried sum rule is directly incorporated 
into the normalization of parton densities. Here we stick to the standard 
approximation, where deutron and carbon parton densities contain only 
singlet parts and a completely nonsinglet density contributes to the difference 
between deutron and proton cases.

Note that the result~(\ref{B}) cannot be directly obtained in the standard calculations of $F_2$ Mellin
moments through the optical theorem and the results for the forward cross-sections
(see, for example, review~\cite{Kotikov:2007ua} and discussions therein).
The standard approach is to calculate only even Mellin moments while the result in~(\ref{B}) 
comes from the analytic continuation~\cite{Kazakov:1987jk,Kotikov:2005gr} of the results for even Mellin moments or through an integration of the corresponding splitting functions in $x$-space,
obtained in their turn from the even Mellin moments
and some symmetric properties~\cite{Ross:1978xk,Kataev:2003en}. Both approaches lead to the same results.

The Gottfried sum rule $I_G$ has the form 
\be
I_G(Q^2) = 1
-2\int_0^1 \frac{dx}{x} \left({\bf f}_{\overline{d}}(x,Q^2)-{\bf f}_{\overline{u}}(x,Q^2)\right),
\label{4.2}
\ee

Experimentally, at $Q^2_c=4$ GeV$^2$ (see~\cite{Arneodo:1994sh}),
\be
I_G(Q^2_c) = 0.705 \pm 0.078,~~  
\label{I_G_exp}
\ee
i.e. it contains quite strong contribution from the second term, i.e. from the nonsymmetric sea.

\subsection{Fitting procedure}
As in all our previous papers on the subject we use the Jacobi polynomial expansion method (see, e.g.
\cite{Parisi:1978jv,Kri,Kataev:1996vu}).
With the QCD expressions for the Mellin moments $M_n^{\rm pQCD}(Q^2)$ (see, for example, \cite{KK2009})
the SF $F_2^{\rm pQCD}(x,Q^2)$ is reconstructed by using the Jacobi polynomial expansion method:
$$
F_{2}^{\rm pQCD}(x,Q^2)=x^a(1-x)^b\sum_{n=0}^{N_{max}}\Theta_n ^{a,b}(x)\sum_{j=0}^{n}c_j^{(n)}(a,b )
M_{j+2}^{\rm pQCD} (Q^2)\,,
\label{2.1}
$$
where $\Theta_n^{a,b}$ are the Jacobi polynomials and $a,b$ are the parameters to fit. A condition
imposed on the latter is the requirement of the minimal error in reconstructing the structure functions.
MINUIT package~\cite{MINUIT} is as usual used to minimize $\chi^2$ value in the description of two variables;
namely, the function $F_2$ itself and its logarithmic ``slope'' $d\ln F_2(x,Q^2)/d\ln\ln(Q^2/\Lambda^2)$.
The twist expansion is thought to be applicable above approximately $Q^2 \sim 1$ GeV$^2$,
hence the global cut $Q^2 \geq 1$ GeV$^2$ imposed throughout.

As the NS analyses results in~\cite{KK2009,Kri} were found to be $N_{max}$ independent (once $N_{max}\ge 8$),
in the present paper we took $N_{max}=8$.

We use free normalization of the data for different experiments. For a reference set, the most stable deuterium BCDMS data at the value of the beam initial energy $E_0=200$ GeV is used. When other datasets are taken as a reference one, variation in the results is found to be negligible.
In the case of the fixed normalization for each and all datasets the fits tend to yield a little bit worse $\chi^2$.

\section{$Q^2$-dependence of SF moments}
$Q^2$-dependence of the twist-2 part of the SF moments
\be
M_n
(Q^2)=\int_0^1 x^{n-2}\,F_2
(x,Q^2)\,dx
\label{1.a}
\ee
has the following form
\be
M_n^{NS}(Q^2) = R_n^{NS} \times \tilde{M}_n^{NS}(Q^2)
\label{3.a}
\ee
where $R_n^{NS}$ is a normalization constant and 
the (perturbatively calculated) factor $\tilde{M}$ contains the product of the coefficient function
$ C_{NS}(n, \ar(Q^2))$ and ``renormgroup exponent'' $h^{NS}(n, Q^2)$
\be 
\tilde{M}_n^{NS}(Q^2) = C_{NS}
(n, \ar(Q^2)) h^{NS}(n, Q^2)
\ee
with
\be
C_{NS}(n, \ar(Q^2)) = 1 
+ \ar(Q^2) B_{\rm{NS}}^{(1)}(n)
+ \ar^2(Q^2) B_{\rm{NS}}^{(2)}(n) + {\cal O}(\ar^3(Q^2))\,,
\label{1.cf}
\ee
and
\be
h^{NS}(n, Q^2)  = \ar(Q^2)^{\frac{\gamma_{NS}^{(0)}(n)}{2\beta_0}}
\left[1 + \ar(Q^2) Z_{\rm{NS}}^{(1)}(n) + \ar^2(Q^2) Z_{\rm{NS}}^{(2)}(n)
+ {\cal O}\left(\ar^3(Q^2)\right)\right] \,,
\ee
where
\begin{eqnarray}
Z_{\rm{NS}}^{(1)}(n) &=& \frac{1}{2\beta_0} \biggl[ \gamma_{NS}^{(1)}(n) -
  \gamma_{NS}^{(0)}(n)\, b_1\biggr]\,,~~~ b_i=\frac{\beta_i}{\beta_0},
\nonumber \\
Z_{\rm{NS}}^{(2)}(n)&=& \frac{1}{4\beta_0}\left[
\gamma^{(2)}_{NS}(n)
-\gamma^{(1)}_{NS}(n)b_1 + \gamma^{(0)}_{NS}(n)(b^2_1-b_2) \right]
+  \frac{1}{2} {\left(Z^{(1)}_{NS}(n)\right)}^2
\,.
\label{3.21}
\end{eqnarray}
Here $\gamma_{NS}^{(k)}(n)$ are the factors in front of $\ar$ in the expansion
with respect to the latter of the anomalous dimensions $\gamma_{NS}(n,\ar)$.

The evolution of moments $M_n(f,Q^2)$ within an interval with the same value of $f$ active quarks has a simple form
\cite{KKPS}
\be
\frac{M_n(f,Q_1^2)}{M_n(f,Q_{2}^2)} =  \frac{\tilde{M}_n(f,Q_{1}^2)}{\tilde{M}_n(f,Q_2^2)} \,
\label{Mn.0}
\ee

\section{Thresholds}
Following~\cite{KKPS}, $F_2(x,Q^2)$~(\ref{1.1}) in the nonsinglet approximation as well as its moments~(\ref{1.a})
at the $f$-quark threshold $Q^2=Q^2_f$ should be smooth ($NS$ symbol is omitted in this section):
\be
M_n(f,Q^2_f)=M_n(f-1,Q^2_f) \, ,
\label{Mn.1}
\ee
where $f$ denotes a $Q^2$ interval with $f$ active quarks.

By comparing Eqs.~(\ref{3.a}) and~(\ref{Mn.1}), the following relation for a normalization constant $R_n(f)$ is deduced:
\be
\frac{R_n(f)}{R_{n}(f-1)} =  \frac{\tilde{M}_n(f-1,Q_{f}^2)}{\tilde{M}_n(f,Q_f^2)} \,.
\label{Mn.1a}
\ee

Thus, if $Q^2$ and $Q^2_0$ values belong to the intervals with, e.g., $f=4$ and $f=5$, respectively, then the evolution of $I_G$ from $Q^2_0$ to $Q^2$ should contain the following factor
\be
\frac{M_n(f=4,Q^2)}{M_n(f=4,Q_{f=5}^2)} \, \frac{M_n(f=5,Q_{f=5}^2)}{M_n(f=5,Q_0^2)} \,
\label{Mn.2}
\ee
which, according to~(\ref{Mn.1}), simplifies to
\be
\frac{M_n(f=4,Q^2)}{M_n(f=5,Q_{0}^2)}= \frac{\tilde{M}_n(f=4,Q^2)}{\tilde{M}_n(f=5,Q_{0}^2)} \,
\frac{R_n(f=4)}{R_{n}(f=5)}
\label{Mn.3}
\ee
and the coupling constant acquires additional terms (see~\cite{Kniehl:2006bg} and references therein)
\bea
&&\frac{a_s(f-1,Q^2_f)}{a_s(f,Q^2_f)}= 1- \frac{2}{3} l_f a_s(f,Q^2_f) + \frac{4}{3} \left( l_f^2 - \frac{57}{2} l_f + 
 \frac{11}{2}\right)  a_s^2(f,Q^2_f)
\label{asf} \\
&&\frac{a_s(f,Q^2_f)}{a_s(f-1,Q^2_f)}= 1+ \frac{2}{3} l_f a_s(f-1,Q^2_f) + \frac{4}{3} \left( l_f^2 + \frac{57}{2} l_f - 
 \frac{11}{2}\right)  a_s^2(f-1,Q^2_f)
\label{asf.1}
\eea
with 
\be
l_f =\ln(Q^2_f/m^2_f)\,,
\label{lf}
\ee
where $m_f$ is the mass of $f$-flavor quark.

Quite the same way, if $Q^2$ and  $Q^2_0$ values belong to the intervals with $f=3$ and $f=5$,
respectively, then the evolution of $M_n$ from $Q^2_0$ to $Q^2$ should contain the following factor
\be
 \frac{M_n(f=3,Q^2)}{M_n(f=3,Q_{f=4}^2)} \,
\frac{M_n(f=4,Q^2_{f=4})}{M_n(f=4,Q_{f=5}^2)} \, \frac{M_n(f=5,Q_{f=5}^2)}{M_n(f=5,Q_0^2)} \, ,
\label{Mn.4}
\ee
which, according to~(\ref{Mn.1}), simplifies to
\be
\frac{M_n(f=3,Q^2)}{M_n(f=5,Q_{0}^2)} = \frac{\tilde{M}_n(f=3,Q^2)}{\tilde{M}_n(f=5,Q_{0}^2)} \,
\frac{R_n(f=3)}{R_{n}(f=5)} \,.
\label{Mn.5}
\ee
In other words, the values of the moments at thresholds do not contribute. They always cancel each other out according to~(\ref{Mn.1})\footnote{This statement is correct exactly; however, 
the normalization constant $R_n(f)$  depends on the values of the moments at thresholds as seen from~(\ref{Mn.1a}).
Consequently, the moments at thresholds contribute through the normalization constant $R_n(f)$
and in certain order of PT there is some difference between them (see, e.g.,~\cite{KKPS} for more detail).
}.

In general, if $Q^2$ ($Q^2_0$) value belongs to the interval with $f$ ($f_0$) active quarks, the evolution
is found to be very simple
\be
\frac{M_n(f,Q^2)}{M_n(f_0,Q_{0}^2)} = \frac{\tilde{M}_n(f,Q^2)}{\tilde{M}_n(f_0,Q_{0}^2)} \,
\frac{R_n(f)}{R_{n}(f_0)} \,.
\label{Mn.6}
\ee


Turning back to the Gottfried sum rule we have that
$I_G(f,Q^2) \equiv M_{n=1}(f,Q^2)$ is smooth at $Q^2=Q^2_f$; therefore,
\be
I_G(f,Q^2_f) = I_G(f-1,Q^2_f)
\label{IGf}
\ee
and the evolution has the form (if $Q^2$ ($Q^2_0$) value belongs to the interval with $f$ ($f_0$) active quarks):
  \be
\frac{I_G(f,Q^2)}{I_G(f_0,Q_{0}^2)} = \frac{\tilde{M}_{n=1}(f,Q^2)}{\tilde{M}_{n=1}(f_0,Q_{0}^2)} \, ,
\frac{R_{n=1}(f)}{R_{n=1}(f_0)} \, .
\ee

Note that the values of $\tilde{M}_{n}$ simplify at $n=1$, since $B_{\rm{NS}}^{(1)}(n=1)=0$ and
$\gamma_{NS}^{(0)}(n=1)=0$.

Indeed,
\be
C_{NS}
(n=1, \ar(Q^2)) = 1 
+ \ar^2(Q^2) B_G
+ {\cal O}(\ar^3(Q^2))\,.
\ee
and
\be
h^{NS}(n, Q^2)  =
1+ a_s(Q^2) d_1 + \frac{1}{2} \, a_s^2(Q^2) \Bigl[ d_2 + \Bigl(d_1-b_1\Bigr)d_1\Bigr],~~ 
b_1=\frac{\beta_1}{\beta_0},~~ d_i=\frac{\gamma_{\rm{NS}}^{(i)}(n=1)}{2\beta_0},
\ee
with \cite{Kataev:1996vu,Kotikov:2005gr}
\be
\gamma_{\rm{NS}}^{(1)}(n=1)=\frac{8}{9}\Bigl[13+8\zeta_3-12\zeta_2 \Bigr] \approx 7.28158,~~ 
\gamma_{\rm{NS}}^{(2)}(n=1) \approx 161.713785 - 2.429260 f \, .
\ee
(see also Eqs. (\ref{I_G}) and  (\ref{B}) given above).

Note that $I_2(Q^2)$ can be obtained at any $Q^2$ by inverting Eq.~(\ref{I_G})
\be
I_2(Q^2) = \frac{I_G(Q^2)}{1+B_G a_s^2(Q^2)},
\label{I_2}
\ee
where $B_G$ is given in Eq.~(\ref{B}).


\section{Results}
As the valence quark analysis does not deal with gluons the cut imposed on the Bjorken variable ($x\geq 0.25$) effectively excludes the region where gluon density is believed to be non-negligible.
Concerning a twist expansion it is applicable only above $Q^2 \sim 1$ GeV$^2$
hence the cut $Q^2 \geq 1$ GeV$^2$ imposed on data throughout.

A starting point of the evolution is $Q^2_0$ = 90 GeV$^2$.
This $Q^2_0$ value is close to the average value of $Q^2$ spanning the respective data.
The heavy quark thresholds are taken at $Q^2_f=m^2_f$  with $m_c=1.27$ GeV and  $m_b=4.28$ (see~\cite{Breview}),
respectively.

\subsection {PDFs, strong coupling constant and high twist correction }

As in~\cite{KK2001,KKPS} the data with largest systematic errors are cut out by imposing certain limits on the kinematic variable $Y$.
The cuts imposed are $x\geq 0.25$ and $N_{Y_{cut}}=5$ (see Table~1 in \cite{KK2001,KKPS}).
Then, a complete set of data consists of 756 points.

The following values for PDF parametrization~(\ref{4})  are obtained in the fits for $Q_0^2=90$ GeV$^2$:\\
in the case of the fixed $\lambda_{\rm{H}}(Q^2) =0.5$
\bea
A(D_2) &=& 2.362 \pm 0.068,~~~\, A(C) ~=~ 3.301 \pm 0.065,\nonumber \\
b(H_2) &=& 4.256 \pm 0.059,~~~\, b(D_2) ~=~ 4.228 \pm 0.022,~~~\, b(C) ~=~ 4.224 \pm 0.041,\nonumber \\
d(H_2) &=& 12.16 \pm 2.35,~~~\,  d(D_2) ~=~ 3.956 \pm 0.258,~~~\,  d(C) ~=~ 1.990 \pm 0.252;\nonumber \\
\eea
in the case of the free $\lambda_{\rm{H}}(Q^2)$ values
\bea
\lambda_{\rm{H}}  &=& 0.742 \pm 0.043,~~~\,
A(D_2) ~=~ 2.362 \pm 0.070,~~~\, A(C) ~=~ 3.294 \pm 0.054,\nonumber \\
b(H_2) &=& 4.314 \pm 0.065,~~~\, b(D_2) ~=~ 4.228 \pm 0.023,~~~\, b(C) ~=~ 4.226 \pm 0.041,\nonumber \\
d(H_2) &=& 3.979 \pm 0.941,~~~\,  d(D_2) ~=~ 3.957 \pm 0.266,~~~\,  d(C) ~=~ 2.001 \pm 0.248.\nonumber
\eea
They are seen to be very similar to those presented in~\cite{KKPS}.


{\bf Table 1.} {\sl Parameter values of the twist-four term $\tilde h_4(x)$ }
\begin{center}
\begin{tabular}{|c||c|c||}
\hline
$x$ & $\lambda_{\rm{H}} =0.5$ & free $\lambda_{\rm{H}}$   \\
\hline \hline
0.275&-0.167$\pm$0.011& -0.167$\pm$0.013\\
0.35 &-0.205$\pm$0.007& -0.205$\pm$0.007\\
0.45 &-0.161$\pm$0.020& -0.162$\pm$0.020\\
0.55 &-0.137$\pm$0.040& -0.138 $\pm$0.040\\
0.65 &-0.129 $\pm$0.072& -0.132 $\pm$0.074\\
0.75 &-0.148 $\pm$0.117& -0.148 $\pm$0.121\\
\hline
\end{tabular}
\end{center}
\vspace{0.5cm}

Twist-4 parameter values are presented in Table~1. 
It is seen that they are almost the same for both cases and, moreover,
are observed to be 
very similar to those given in~\cite{KKPS}. 

Finally, using the nonsinglet evolution analyses of SLAC, NMC and BCDMS experimental data for SF $F_2$ 
for the fixed $\lambda_{\rm{H}} =0.5$
we obtain (with $\chi^2/DOF=1.03$)
\bea
\as(M_Z^2) ~=~ 0.11795 +
\biggl\{ \begin{array}{l} \pm 0.0004 ~\mbox{(stat)}
\pm 0.0018~\mbox{(syst)}  \pm 0.0006 ~\mbox{(norm)}\\
\pm 0.0019~\mbox{(total exp.error)}
 \end{array} \,.
\label{la_05}
\eea

With free $\lambda_{\rm{H}}(Q^2)$,
the following result is obtained (with $\chi^2/DOF=0.88$):
\bea
\as(M_Z^2) ~=~ 0.11798 +
\biggl\{ \begin{array}{l}
\pm 0.0003 ~\mbox{(stat)}
\pm 0.0019 ~\mbox{(syst)}  \pm 0.0005 ~\mbox{(norm)} \\
\pm 0.0020 ~\mbox{(total exp.error)}
\end{array} \,.
\label{free_la}
\eea

It is seen that the values of the strong coupling constant $\as(M_Z^2)$ are nearly the same in the cases of fixed and free values of $\lambda_{\rm{H}}$.

\subsection {Second moment}
The second moment $n=2$ of the difference in valence parts of $u$ and $d$ quark distributions is also investigated in the lattice models. Following~\cite{ABM,Blumlein:2006be}, we will estimate this second moment illustrating some features of the nucleon structure.

This difference in valence $u$ and $d$ quark distributions can be extracted at large $x$ values directly from that of the nonsiglet parton densities in proton and deuteron
\be
 \tilde{\bf f}_u^v(x,Q^2) - \tilde{\bf f}_d^v(x,Q^2) \approx  \tilde{\bf f}_{\rm{H}}(x,Q^2)-\tilde{\bf f}_{\rm{D}}(x,Q^2),
\label{diff}
\ee
since a contribution of the sea quarks and antiquarks are negligible here.
Indeed, in agreement with the quark counting rules~\cite{Matveev:1973ra}, at large $x$ values 
sea quark density is suppressed by the additional factors $\sim (1-x)$ and $\sim (1-x)^2$ as compared to gluon
and valence quark ones, respectively.
It is seen that the contribution coming from sea quarks and antiquarks is even less important than that caused by the gluons,
and both these contributions are negligible at large $x$ values.

Thus, using results given in~(\ref{la_05}) and~(\ref{free_la})
the respective difference for the second moments ${\bf f}_u^v(2,Q^2)-{\bf f}_d^v(2,Q^2)$ is shown in Table~2.

{\bf Table 2.} The difference ${\bf f}_u^v(2,Q^2)-{\bf f}_d^v(2,Q^2)$.
\vspace{0.2cm}
\begin{center}
\begin{tabular}{|c|c|c|c|c|}
\hline
        &               &      &                               &         \\
$Q^2$, GeV$^2$ & 90 & 4 & 2 & 1  \\
        &               &      &                               &          \\ \hline \hline
 $\lambda_{\rm{H}}=0.5$     &    0.110$\pm$0.012 &  0.139$\pm$0.016  & 0.150$\pm$0.017  & 0.179$\pm$0.020 \\
free $\lambda_{\rm{H}}$     &    0.115$\pm$0.013 &  0.145$\pm$0.016  & 0.157$\pm$0.017  & 0.187 $\pm$0.021 \\
\hline
\end{tabular}
\end{center}

Following~\cite{Kotikov:2014cua}, we would like to note that the lattice results are strongly 
nonperturbative and, therefore, it is better to compare them with our result obtained at $Q^2$= 1 GeV$^2$, which corresponds in the present analysis to the boundary between
perturbative and nonperturbative QCD. A comparison of our result (at $Q^2$= 1 GeV$^2$) with those
derived in the lattice QCD approach is shown in Fig.~1.
The  lattice points are taken from the recent paper~\cite{Abdel-Rehim:2015owa}.

It is seen that the result of the analysis with free $\lambda_{\rm{H}}$ almost coincides with that presented in~\cite{Kotikov:2014cua}.
However, results presented here demonstrate considerably smaller uncertainties, which is a direct consequence of the new form (\ref{4}) and (\ref{4.1}) adopted in the PDF parametrizations.

\begin{figure}[!htb]
\unitlength=1mm
\vskip -1.5cm
\begin{picture}(0,100)
  \put(0,-5){%
   \psfig{file=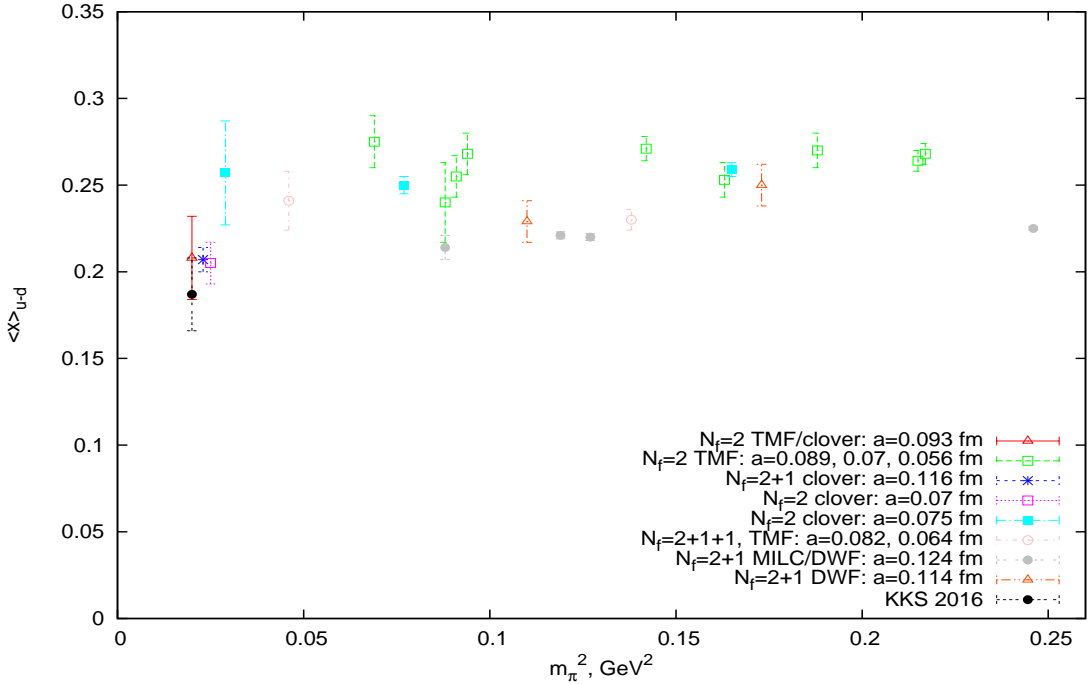,width=150mm,height=95mm}
}
\end{picture}
\vskip 0.2cm
\caption{ Lattice computations results for the second moment (or isovector nucleon momentum fraction)
vs a pion mass $m_{\pi}$ as borrowed from~\cite{Abdel-Rehim:2015owa}; our KKS 2016 point for 
$<\!\!x\!\!>_{u-d} \equiv {\bf f}_u^v(2,Q^2) - {\bf f}_d^v(2,Q^2)$ is obtained at $Q^2$= 1 GeV$^2$.}
\end{figure}

\section{Conclusions}
In this work the Jacobi polynomial expansion method developed in~\cite{Parisi:1978jv,Kri} was used to 
analyze $Q^2$-evolution of DIS structure function $F_2$
by fitting reliable fixed-target experimental data that satisfy the cut $x \geq 0.25$.
Based on the results of fitting, the strong coupling constant is evaluated at the normalization point.

The present results are compatible with those obtained in our previous papers~\cite{KKPS,Kotikov:2014cua}. 
Moreover, they almost coincide for both cases($\lambda_{\rm{H}}=0.5$ fixed and free $\lambda_{\rm{H}}$ one). The only difference is in the uncertainties. 
For $\lambda_{\rm{H}}=0.5$ we have
\be
\as(M_Z^2) = 0.1180 \pm 0.0019~\mbox{(total exp.error)}, \label{re1n} \\
\ee
With freely varied values of $\lambda_{\rm{H}}$
we obtain 
\be
\as(M_Z^2) = 0.1180 \pm 0.0020~\mbox{(total exp.error)} \, . \label{re2n} \\
\ee

Our result almost coincides with the central world average value
\be
\as(M_Z^2)|_{\rm{world~average}} = 0.1181 \pm 0.0011 \, ,
\label{world} \\
\ee
presented in~\cite{Breview}.

Good agreement with a recent result obtained in lattice QCD~\cite{Abdel-Rehim:2015owa} is also observed, which is exhibited by the results for the second moment $<\!\!x\!\!>_{u-d}$ (see Fig.~1).

As the next steps of our investigations, we plan to perform  N$^3$LO fits at large $x$ values.
In order to do that we use the contributions
coming from three loops in the coefficient functions~\cite{Vermaseren:2005qc}, as well as
four-loop corrections to the first several moments of anomalous dimensions (see~\cite{Velizhanin:2014fua,Baikov}
and discussions therein). The knowledge of first several moments of anomalous dimensions is enough to do analyses of such a type. In a sense it is like the first NNLO analysis performed in~\cite{PKK}.
We note that several N$^3$LO fits had already been done in~\cite{Kataev:2001kk,Blumlein:2006be} with suggestion about a negligible contributions coming from the four-loop anomalous dimensions.

We plan to apply some resummations, like a Grunberg effective charge method~\cite{Grunberg:1980ja}
(as it was done in~\cite{Kotikov:1992ht} at the NLO approximation) and the ``frozen'' and analytic modifications of the strong coupling constant (see~\cite{Kotikov:2010bm} and references therein).
Note that the resummations and the infrared safe modifications of 
the strong coupling constant often lead to the agreement between experimental data and theoretical predictions significantly improved (see~\cite{Kotikov:1992ht,Kotikov:1993yw} and \cite{Kotikov:2010bm,Cvetic:2009kw}, respectively, and references and discussions therein).
It is hoped that similar properties will be observed in the forthcoming investigations.

\section{Acknowledgments}
The work was supported by RFBR grant No.16-02-00790-a.

\end{document}